\documentclass[aps,pre,twocolumn,a4paper,floatfix,showpacs]{revtex4-1}
\usepackage{hyperref}
\usepackage{natbib}
\usepackage{graphicx}
\usepackage{color}
\usepackage{amsmath}
\usepackage[capitalise]{cleveref}
\begin{document}
\title{Damping filter method for obtaining spatially localized exact solutions}
\author{Toshiki Teramura}
\email{teramura@kyoryu.scphys.kyoto-u.ac.jp}
\affiliation{Department of Physics and Astronomy, Graduate School of Science, Kyoto University, Japan}
\author{Sadayoshi Toh}
\affiliation{Department of Physics and Astronomy, Graduate School of Science, Kyoto University, Japan}
\date{\today}
\begin{abstract}
Spatially localized structures are key components of turbulence and other spatio-temporally chaotic systems.
From a dynamical systems viewpoint, it is desirable to obtain corresponding exact solutions,
though their existence is not guaranteed.
A damping filter method is introduced to obtain variously localized solutions,
and adopted into two typical cases.
This method introduces a spatially selective damping effect to make a good guess at the exact solution,
and we can obtain an exact solution through a continuation with the damping amplitude.
First target is a steady solution to Swift-Hohenberg equation,
which is a representative of bi-stable systems in which localized solutions coexist,
and a model for span-wisely localized cases.
Not only solutions belonging to the well-known snaking branches
but also those belonging to an isolated branch known as ``isolas'' are found
with a continuation paths between them in phase space extended with the damping amplitude.
This indicates that this spatially selective excitation mechanism has an advantage
in searching spatially localized solutions.
Second target is a spatially localized traveling-wave solution to Kuramoto-Sivashinsky equation,
which is a model for stream-wisely localized cases.
Since the spatially selective damping effect breaks Galilean and translational invariances,
the propagation velocity cannot be determined uniquely while the damping is active,
and a singularity arises when these invariances are recovered.
We demonstrate that this singularity can be avoided by imposing a simple condition,
and a localized traveling-wave solution is obtained with a specific propagation speed.
\end{abstract}
\maketitle
\section{introduction}
A dynamical systems point of view and accompanying exact solutions to Navier-Stokes equation
play key roles in understanding the dynamics of turbulence \cite{Kawahara2012}.
It is useful both in transient flows and statistically steady flows.
Indeed in minimal turbulence \cite{Jimenez1991},
a crucial aspect of the dynamics of laminar-turbulent transition processes has been revealed by
the discovery of ``edge state'' \cite{JPSJ.70.703,PhysRevLett.96.174101},
and that of the self-sustaining process (SSP) has been done by
the discovery of unstable periodic orbits which reproduce the statistics of turbulence
\cite{CambridgeJournals:91341}.

The dynamical systems viewpoint, however, has not yet successfully captured
the full nonlinear spatio-temporal dynamics of turbulence.
One major limitation is that there is no general framework for obtaining
solutions corresponding to spatially localized structures in turbulence.
For example, in channel flows there exist
near-wall structures and large scale motion (LSM)
that occupies the outer layer above the near-wall layer.
In order to elucidate their intrinsic dynamics and interactions among them
from the dynamical systems viewpoint,
it is desireble to obtain corresponding exact solutions separately.
Such localized exact solutions are paid much attention recent years,
and indeed obtained in a few cases
\cite{PhysRevLett.99.034502,duguet:111701,CambridgeJournals:7324528,Avila2013}.
However, despite of their importance,
any practical ways to obtain them have not established yet.
For this purpose, we introduce a damping filter method in \cref{sec:Method}.

There exist various types of localized structures.
For example, turbulent puffs in pipe flow are localized in the stream-wise direction;
turbulent spots in channel flow do both in the stream-wise and span-wise directions.
If there exists a solution corresponding to LSM,
it will be localized in the wall-normal direction.
In this paper, we focus on the two typical cases both of which will be instructive
in understanding localized structures observed in turbulence.
We show not only basic usage and results but also remarkable features of
our method in \cref{sec:SHE,sec:KSE}.

We first consider localized solutions to Swift-Hohenberg equation (SHE)
\cite{burke2007homoclinic,Knobloch2008,Beck2009}.
This is a representative example
of localized solutions in bi-stable systems.
This class of localized solutions contains,
for example, span-wisely localized solutions corresponding to roll-streak structures
in plane Couette flow \cite{Schneider2010a}.
Their solution branches are very similar
to the ``snaking'' branches seen in SHE \cite{burke2007homoclinic}.
Similar structures of solution branches are also found
in doubly diffusive convection systems \cite{Bergeon2008}.
These facts indicate that there exists a universal mechanism
of spatially localized solutions in the bi-stable systems.
We deal with solutions in this class in \cref{sec:SHE}.

Second, we examine a spatially localized traveling-wave solution
to Kuramoto-Sivashinsky equation.
Such a solution can be regarded as a stream-wisely localized solution.
Stream-wisely localized structures can be observed in pipe flow (turbulent puffs),
boundary layers (hairpin vortexes), and so on.
The sustaining mechanism of them might be different
from that of span-wisely localized solutions,
and thus it is necessary to obtain the corresponding solutions
in order to analyze them from a dynamical systems viewpoint.
At a glance, since our method utilizes a spatially selective damping effect
that breaks translational invariance,
it seems to have only limited capability for this issue.
However, we show that this is not the case in \cref{sec:KSE}.

This paper is organized as follows.
In \cref{sec:Method},
a damping filter method is introduced, where we explain
its concept and concrete procedure.
In \cref{sec:SHE,sec:KSE}
we adopt the method to Swift-Hohenberg equation
and Kuramoto-Sivashinsky equation respectively
in order to obtain spatially localized solutions.
Finally, this paper is concluded with concluding remarks in \cref{sec:ConcludingRemarks}.

\section{\label{sec:Method}damping filter method}
In this section the protocol of the damping filter method is explained.
This method consists of three steps.
\begin{figure}[btp]
    \includegraphics[width=6.5cm]{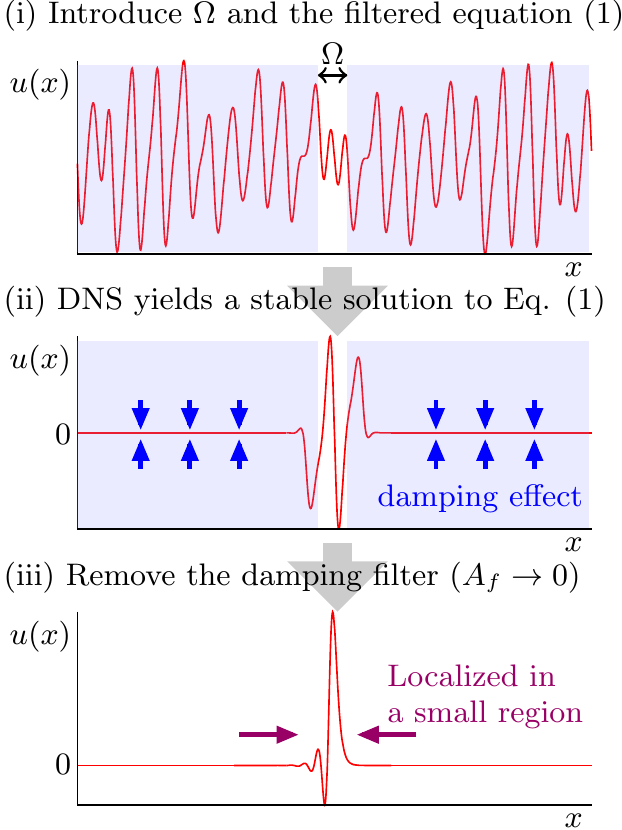}
    \caption{
        (Color online)
        Three steps in the damping filtering method.
    }
    \label{fig:MainIdeas}
\end{figure}
Since this method can be used for various systems,
a general form evolution equation, $\partial_t u = F[u]$,
is used in the following explanation.

The first step of this method is to introduce
a spatially selective damping term into the evolution equation.
The damping term works only in a region $\Omega$.
If we want to obtain a span-wisely localized solution,
$\Omega$ should be a region localized in span-wise direction.
If we want to obtain a solution localized like turbulent spot,
$\Omega$ should be a spot region.
Then the damping term is introduced as follows:
\begin{equation}
    \partial_t u(x,t) = F[u](x,t) - A_f H(x) u(x,t).
    \label{eq:FilteredEquation}
\end{equation}
We call this equation ``filtered equation'' hereafter.
The damping term consists of the filter amplitude $A_f$ and a filter function $H(x)$.
$H(x)$ is defined to be zero in the region $\Omega$ and to be $1$ out of $\Omega$,
and smoothed by taking a convolution with a mean-zero Gaussian $\mathcal{N}_{0,\sigma^2}(x)$
in order to avoid the numerical singularity:
\begin{eqnarray}
    H(x) = \int \mathcal{N}_{0,\sigma^2}(x-y) \hat{H}(y) dy, \\
    \hat{H}(x) = \begin{cases} 0 & (x \in \Omega) \\ 1 & (x \not\in \Omega) \end{cases}.
    \label{eq:FilterFunction}
\end{eqnarray}
The integral is taken in the whole region.
The damping term causes a linear damping effect out of $\Omega$ (filtered region),
and the filtered equation equals to the original equation
$\partial_t u = F[u]$ in $\Omega$ (unfiltered region).

The second step is to obtain an exact solution to the filtered equation (\ref{eq:FilteredEquation}).
Owing to the spatially selective damping effect,
the direct numerical simulation (DNS) of the filtered equation
tends to yield a spatially localized time series $u(x,t)$,
in other words, $u(x,t)$ decreases exponentially fast as $x$ goes away from $\Omega$
after a relaxation time.
In addition, since the damping effect weakens the instability of the system,
DNS sometimes yields a non-trivial stable solution
for enough large filter amplitudes and an appropriate $\Omega$.
In this case the second step is finished with this stable solution.
If any stable solutions are not obtained,
a solution to the filtered equation is obtained
by solving an equation $F[u] - A_f H(x) u = 0$ about $u(x)$ with Newton method.
Since the degree of freedom is also reduced by the damping term,
it is expected that such a solution can be obtained easily.

The third step is a continuation process.
The solution obtained in the second step depends on the filter amplitude $A_f$,
and often this dependency is continuous.
A continuation with $A_f$ is started from this solution.
The filter amplitude $A_f$ is decreased until it gets to zero,
where the filtered equation is restored to the original equation in the whole region.
Then the continuated solution is nothing but that to the original equation.
This is the goal of the damping filter method.
The continuation is implemented by the arc-length method
with Newton-Krylov iterative method,
and thus applicable to systems having large degree of freedom.

The good feature of our method is that
an appropriate guess of a spatially localized solution is constructed
as a solution to the filtered equation (\ref{eq:FilteredEquation}).
This guess reflects the dynamics of spatially localized structures
since the filtered equation equals to the original equation in the region $\Omega$.
In another study \cite{Nagata1990},
an artificial external forcing is used for constructing a guess of solutions.
It was designed by hand from the inference about the dynamics of localized structures.
In our method such artificial manipulation is not needed
except for determining the region $\Omega$.

This spatially selective damping is inspired
by the work \cite{Jimenez2001}.
They have investigated an autonomous behaviors of
near-wall structures by a filtered dynamics.
In contrast to them,
our method uses this filtered dynamics only for guesses and continuations,
and removes the filter finally.
Thus, our method enable us to study the non-filtered dynamics by localized solutions.

\section{\label{sec:SHE}Span-wisely localized solutions}
In this section we consider one-dimensional Swift-Hohenberg equation (SHE):
\begin{equation}
    \frac{\partial u}{\partial t} = F[u] 
    = \left( r - \left( \frac{\partial^2}{\partial x^2} + 1 \right)^2 \right) u + 2u^3 - u^5.
    \label{eq:SHE}
\end{equation}
As noted in the introduction,
a series of solutions to SHE is regarded as a representative
of localized solutions in the bi-stable systems.
The following subsections show two things:
(i) We can obtain span-wisely localized solutions by our method.
In order to show this,
we reproduce solutions belonging to the homoclinic snaking branches.
The practical detail of our method is also described.
(ii) Our method has a capability for obtaining various solutions
that are usually hard to be found.
Indeed, we find an isolated and closed solution branch.
Since isolated solution branches cannot be found
by the weakly nonlinear framework,
this success indicates an advantage of our method.

\subsection{Homoclinic snaking branches}
In this subsection we apply our method to SHE
in order to obtain a localized steady solution belonging to the snaking branches.
Although the branches contain stable localized solutions for a parameter region,
the attracting basins of them are very small,
and thus it is almost impossible to obtain these localized solutions
by DNSs with arbitrary initial conditions.

Before adopting our method,
the setting of system is described.
We consider SHE in a region $[0,L], L = 180$,
and impose a fixed boundary conditions $u(0) = u(L) = 0$.
The parameter $r$ is set to $-0.669$ in this subsection,
and $-0.633$ is used in the next subsection.
Time evolutions are solved by the quasi-spectral method
with the classical fourth-order Runge-Kutta method.
We regard a steady point of DNS as a steady solution to the equation,
so the DNS code is used also in continuation processes.

We describe the practical details of our method hereafter.
The first step of our method is to introduce the damping term.
The unfiltered region $\Omega$ is set to be $[80,100]$,
and the filter function $H(x)$ is smoothed with $\sigma^2 = 0.01$.
The amplitude of the filter $A_f$ is set to be $1$.

The second step is to obtain an solution
to the filtered equation (\ref{eq:FilteredEquation}) with $F[u]$ of \cref{eq:SHE}.
This equation has a stable localized steady solution with these parameters.
The initial condition of this DNS
is the stable steady sine-like solution of non-filtered equation (\ref{eq:SHE}),
which is not spatially localized but spatially extended.
Such a localized solution to the filtered equation exists while $r \lesssim -0.72$.
This lower limit almost agrees with that of the snaking branches.
Since the spatial period of the sine-like solution is $2\pi$,
these localized solutions to the filtered equation
contain almost three periodic components.

\begin{figure}[tbp]
    \begin{center}
        \includegraphics{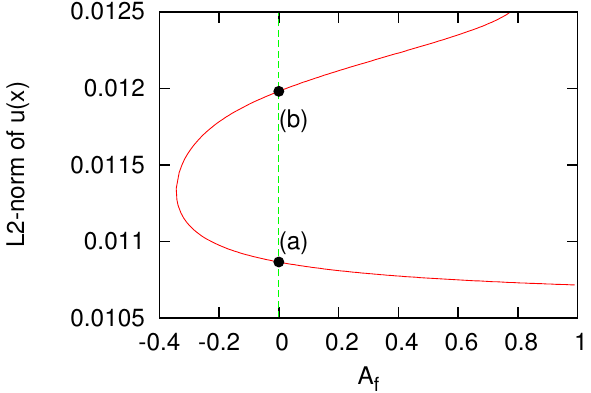}
    \end{center}
    \caption{
        (Color online)
        The trajectory of continuation projected onto a $A_f$-$\| u \|$ space.
        Each of two labeled solutions (a),(b) on the line $A_f = 0$
        denotes a solution to SHE.
        The continuation is continued
        after the filter amplitude $A_f$ became zero,
        and yield a solution labeled (b).
        The profile of these solutions are shown in \cref{fig:SHE_root1}.
    }
    \label{fig:SHE_trace1}
\end{figure}

\begin{figure}[tbp]
    \begin{center}
        \includegraphics{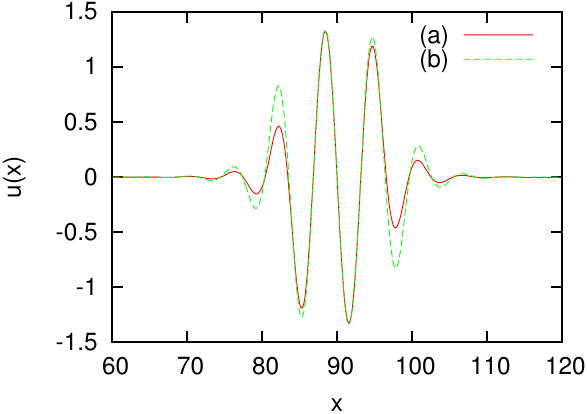}
    \end{center}
    \caption{
        (Color online)
        The profiles of the solutions obtained in the continuation shown in \cref{fig:SHE_trace1}.
        Since the tails of the solutions decay exponentially,
        These solutions are localized almost in $[70,110]$,
        and have an exponentially decaying tail.
    }
    \label{fig:SHE_root1}
\end{figure}

The third step is a continuation process.
The parameter traced in this continuation is the filter amplitude $A_f$,
and the parameter $r$ is fixed.
The result of the continuation is shown in \cref{fig:SHE_trace1},
which displays the trajectory of the continuation projected onto a $A_f$-$\| u \|$ plane.
Here $\| \cdot \|$ denotes the $\text{L}^2$-norm.
The trajectory crosses the line $A_f = 0$ twice.
Although a solution to SHE is obtained when the trajectory crosses the line first
and thus the damping filter method finishes at this time,
we find that the trajectory turns back and crosses the line $A_f = 0$ again.
Eventually, we successfully obtain two solutions to SHE,
and the profile of them are shown in \cref{fig:SHE_root1}.
These solutions are localized almost in $[70,110]$,
which is larger than $\Omega = [80,100]$.
This fact indicates that
$\Omega$ is only a guide for obtaining a spatially localized solution,
which obeys not the damping filter but the original equation.

\subsection{An isolated closed branch}
We execute the same procedure for various values of the parameter $r$.
For most of $r$
it yields continuation trajectories and solutions to SHE
similar to those shown in the previous section.
However, we also find quite different behaviors in some cases,
one of which we focus on in this subsection.

As an initial guess
 we use a solution to the filtered equation
obtained by a continuation with the parameter $r$ started from the solution 
to the filtered equation used in the previous section.
The other parameters are same as those of the previous section: 
$\Omega = [80,100]$ and $A_f = 1$.

\begin{figure}[tbp]
    \centering
    \includegraphics{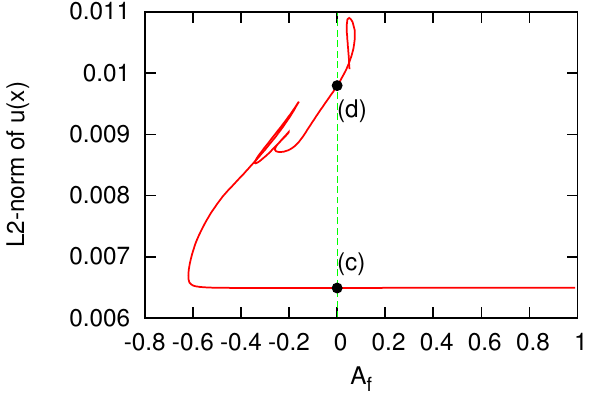}
    \caption{
        (Color online)
        The trajectory of continuation projected onto $A_f$-$\| u\|$
        started from an unstable solution to the filtered equation.
        This continuation also cross the line $A_f = 0$ twice,
        but a complicated path is realized.
        Two labeled points (c) and (d) on the line $A_f=0$
        are also solutions to SHE,
        and their profiles are shown in \cref{fig:SHE_root2}.
        As noted in the text
        a switching between solution branches are occured
        in this complicated part of the trajectory.
    }
    \label{fig:SHE_trace2}
\end{figure}

\begin{figure}[tbp]
    \centering
    \includegraphics{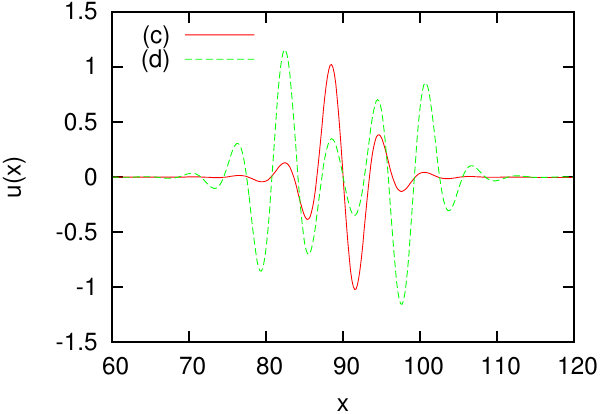}
    \caption{
        (Color online)
        The profiles of the solutions (c) and (d).
        In contrast to \cref{fig:SHE_root1},
        the profiles of the solutions (c) and (d) are qualitatively different.
    }
    \label{fig:SHE_root2}
\end{figure}

The result of the continuation with $A_f$ starting from this starting solution
is displayed in \cref{fig:SHE_trace2}.
Compared with \cref{fig:SHE_trace1},
the continuation trajectory is very complicated
especially in the region $A_f < 0$.
The two solutions on the line $A_f = 0$ are labeled as (c) and (d),
and their profiles are displayed in \cref{fig:SHE_root2}.

\begin{figure}[tbp]
    \centering
    \includegraphics{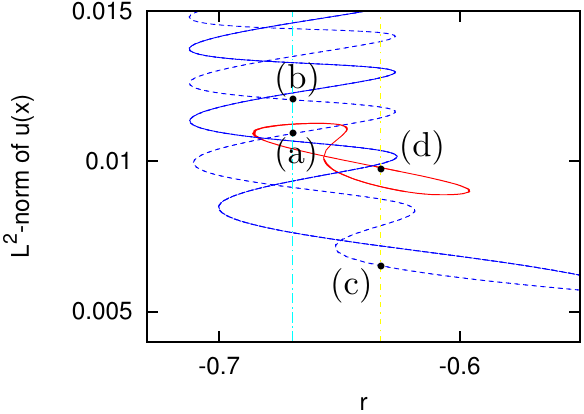}
    \caption{
        (Color online)
        The snaking branches and isolas projected onto a $r$-$\|u\|$ plane.
        The snaking branches consist of two coupled branches,
        and solutions (a), (b) and (c) belong to the same one of them
        (dashed blue line).
        The red (dark gray) figure-eight shaped line
        denotes the isolas containing the solution (d).
        Such eight-figured branches can be seen also in two-pulse solutions to SHE
        \cite{springerlink:10.1007_s10884-010-9195-9}.
        This similarity tells that
        the solution (d) should be regarded as a connected two-pulse solution.
    }
    \label{fig:branches}
\end{figure}

A difference between the solutions (c) and (d) can be observed in its profile:
The solution (c) is a single pulse solution like the solutions (a) and (b).
On the other hand,
the solution (d) seems to be a combination
of two antisymmetric pulse solutions.
There exists the definitive difference
in their solution branches as shown in \cref{fig:branches}.
This figure shows that
the solutions (a), (b), and (c) belong to the snaking branches,
but the solution (d) belongs to an isolated closed branch.
Such closed isolated solution branches are called ``isolas''
in \cite{springerlink:10.1007_s10884-010-9195-9}.

It should be noted that
two distinct branches are connected by the continuation  with $A_f$.
Moreover, they connect through the region where $A_f$ becomes negative,
where the term $-A_f H(x) u$ works as an excitation term.
Thus the connection can be regarded as a result of the instability caused by this term.
Since $u(x)$ has an exponentially decaying tail,
this instability occurs only around the edge of $\Omega$.
So this linear exitation is also spatially selective.

\section{\label{sec:KSE}Stream-wisely localized solutions}
In this section a stream-wisely localized solution,
in other words, a solution localized in  its moving direction is studied
with Kuramoto-Sivashinsky equation (KSE):
\begin{equation}
    \frac{\partial u}{\partial t} = F[u] = - u \frac{\partial u}{\partial x}
    - \frac{\partial^2 u}{\partial x^2} - \frac{\partial^4 u}{\partial x^4}.
    \label{eq:KSE}
\end{equation}
It should be noted that
KSE has no localized equilibrium solution whose tail decays exponentially.
If we adopt our method to \cref{eq:KSE},
the continuation about $A_f$ yields the flat solution $u=0$ before $A_f$ reaches zero.
So we seek a stream-wisely localized traveling-wave solution (TWS)
such that $u(x,t) = \hat{u}_0(x-ct)$ satisfying  boundary conditions
$\hat{u}_0(x-ct \to \pm \infty) \to 0$.

In contrast to the case of span-wisely localized solutions,
there are two issues in this case:
One is a treatment of the propagation velocity of the solution,
and the other is a breakdown of the localization.
These issues will arise in more general cases
since they are based on Galilean and translational invariances.
In the following subsections a solution to these issues are described.

\subsection{A treatment of the propagation velocity}
Since a TWS travels downstream,
its localized region must accompany.
In order to obtain a localized TWS by our method,
it is necessary to introduce a moving damping filter
or a steady damping filter in a moving frame, and we chose the latter.
Then the equation becomes as follows:
\begin{equation}
        \frac{\partial \hat{u}}{\partial t^\prime} =
        - (\hat{u}-c)\frac{\partial\hat{u}}{\partial x^\prime}
        - \frac{\partial^2 \hat{u}}{\partial x^{\prime 2}}
        - \frac{\partial^4 \hat{u}}{\partial x^{\prime 4}}
        - A_f H(x^\prime) \hat{u},
    \label{eq:KSEtw}
\end{equation}
where $x^\prime = x -ct, t^\prime = t, \hat{u}(x^\prime,t^\prime) = u(x,t)$.
We seek a steady solution satisfying $\partial_{t^\prime} \hat{u} = 0$ in this frame.
However, the velocity of the moving frame $c$ is unknown unless the solution is obtained.
Since KSE does not allow their solutions to  continuously depend on $c$,
a specific value $c_0$ with which a solution $\hat{u}_0(x-c_0t)$ exists must be determined
simultaneously.

Such a situation sometimes occurs when obtaining a TWS to an equation $\partial_t u = F[u]$.
In these cases, this problem is usually resolved by regarding $c$ as an unknown variable,
and solving $-c\partial_x{u} = F[u]$ for $u(x)$ and $c$.
Since the translational invariance reduces the degree of freedom,
this simultaneous equation can be solved.
In our method, however,
the damping filter breaks Galilean and translational invariances,
thus the usual technique cannot be adopted.

\begin{figure}[tbp]
    \centering
    \includegraphics{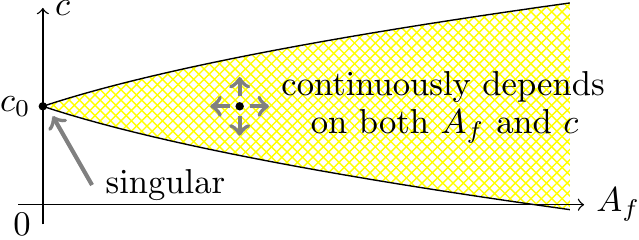}
    \caption{
        (Color online)
        An image for the continuous existence of solutions and singularity at $A_f = 0$.
        The existence of the solution is displayed in $A_f$-$c$ plane.
        The solution exists continuously on the yellow (thin gray) crosshatched region,
        and cannot be traced beyond its rim.
        There is no guarantee that this solution branch connects to the line $A_f=0$. If it does, only a point is allowed
        because the solution cannot exist continuously on the line $A_f = 0$.
    }
    \label{fig:ContiuousExsitance}
\end{figure}

Although TWSs to KSE do not continuously depend on $c$,
our results show that TWSs to the filtered equation do.
This fact reflects the breaking of the translational invariance by the filter.
The details are discussed in the last part of this section.
Thus the propagation velocity $c$ becomes one of the control parameters of the solution and 
can be chosen arbitrarily in a certain range.
See also schematic view in Fig.\ref{fig:ContiuousExsitance}.

Then the other issue arises;
how do we obtain the specific value $c_0$?
Although there is a range of $c$ in which the solution exists continuously,
this range becomes narrower and narrower as $A_f$ goes to zero,
and finally converges to a point on $A_f = 0$.
We get over this issue by imposing an implicit relationship between $c$ and $A_f$.
This technique to impose the restriction is a key point
to obtain a stream-wisely localized TWS.
The details are discussed in the next subsection with our data.

\subsection{Adopting the damping filtering method to KSE}
In the first step of the damping filtering method,
a localized region $\Omega$ is determined.
We set the system size to $L = 200$ and  the localized region $\Omega = [97,103]$
in order to obtain a one-peak TWS.
The filter function $H(x)$ is smoothed by Gaussian with  $\sigma^2 = 0.01$
and the filter amplitude $A_f$ is set to $4.8$.

In the second step, a solution to the filtered \cref{eq:FilteredEquation} is obtained 
to start the continuation.
As notated in the previous subsection,
we can choose an arbitrary propagation velocity $c$ in a certain range,
and we choose $c=0$ here.
We execute a DNS of non-filtered equation (\ref{eq:KSEtw})
to produce a spatio-temporal chaotic field,
which is used for an initial condition of a DNS of \cref{eq:KSEtw}.
This yields a stable solution, which is labeled as (a).
These DNSs are solved by the quasi-spectral method with
the classical forth-order Runge-Kutta method.
Although a localized solution has an exponentially decaying tail 
and does not get exactly to zero in a finite distance,
we regard small values comparable to the truncation error as zero
and assume that the fixed boundary conditions $u(0) = u(L) = 0$ are satisfied.
We use sine transform to ensure this condition with $N=2048$ modes.
\begin{figure}[tbp]
    \centering
    \includegraphics[width=6cm]{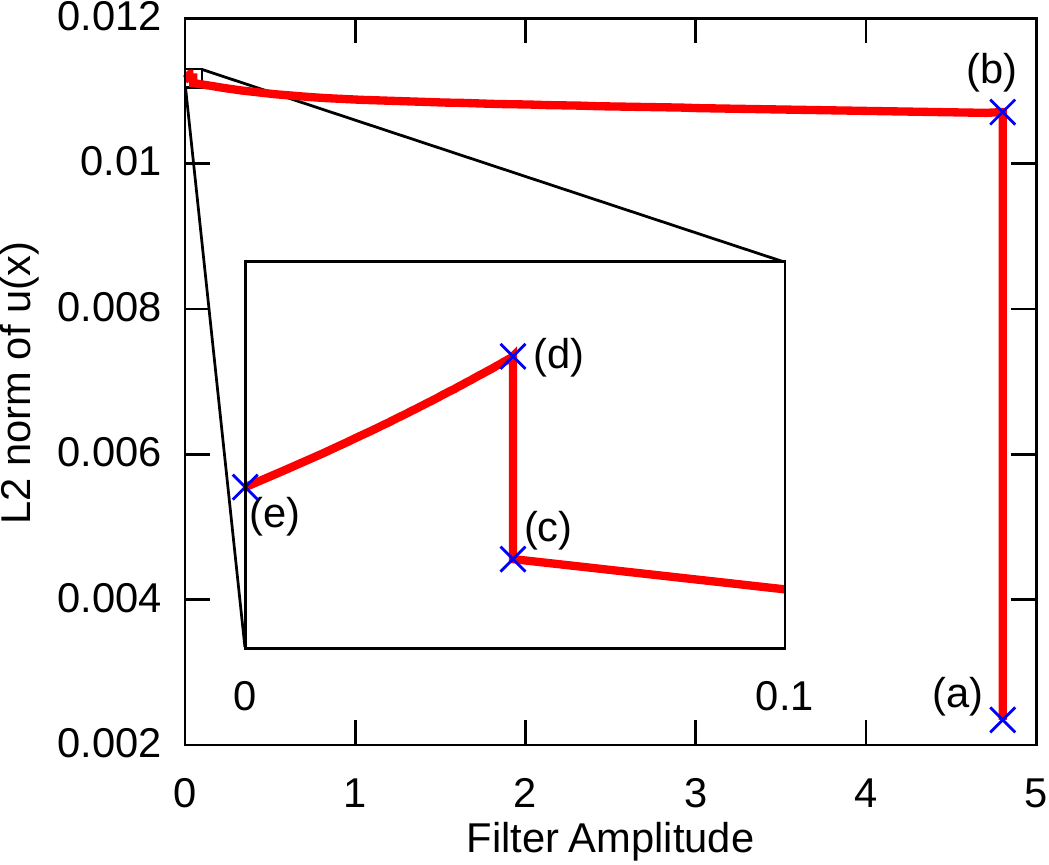}
    \caption{
        (Color online)
        The trajectory of the continuation projected onto $A_f$-$\|u\|$ plane.
        A closeup around $A_f=0$ is also shown.
    }
    \label{fig:kse_tw_trace}
\end{figure}

The third step is a continuation process.
Since the propagation velocity $c$ becomes a continuous parameter
of the solution on $A_f > 0$ region,
this continuation becomes essentially two-dimensional.
It is almost impossible to obtain a full two-dimensional solution branch
because of the numerical cost, we introduce a path on $A_f$-$c$ space as follows.

As noted above,
if $c$ is fixed to zero then a continuation with $A_f$
yields the flat solution $u=0$ before $A_f$ reaches zero.
In order to avoid this dead end,
we first fix $A_f$ to $4.8$ and execute a continuation with $c$.
The solution is traced up to $c \simeq 1.21$, which is labeled as (b).
This value of $c$ is determined by trial and error here,
but a more practical criterion will be discussed in a future work.

\begin{figure}[tbp]
    \centering
    \includegraphics{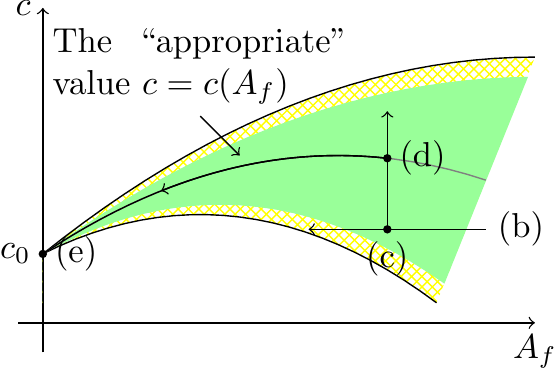}
    \caption{
        (Color online)
        An schematic view of the parameter dependency of the solution around $A_f = 0$.
        The yellow (thin gray) crosshatched region denotes one where a solution continuously exists,
        and the green (thick gray) region denotes one where a ``localized'' solution continuously exists.
        Here a ``localized'' solution means that
        it does not have the oscillation around the boundary.
        The three steps are as follows:
        (i): (b) $\to$ (c),
        (ii): (c) $\to$ (d),
        (iii): (d) $\to$ (e).
    }
    \label{fig:RestrictedPath}
\end{figure}

The following continuation procedure is delicate
because the uniqueness of the propagation velocity $c$ 
must recover when $A_f = 0$.
In short, the procedure consists of three steps
(see \cref{fig:RestrictedPath}):
(i) $A_f$ is reduced around $0.05$ while $c$ is fixed.
(ii) $A_f$ is fixed and $c$ is adjusted to an ``appropriate'' value $c=c(A_f)$.
(iii) $c$ and $A_f$ are traced simultaneously keeping the ``appropriate'' condition $c=c(A_f)$.

\begin{figure}[tbp]
    \centering
    \includegraphics{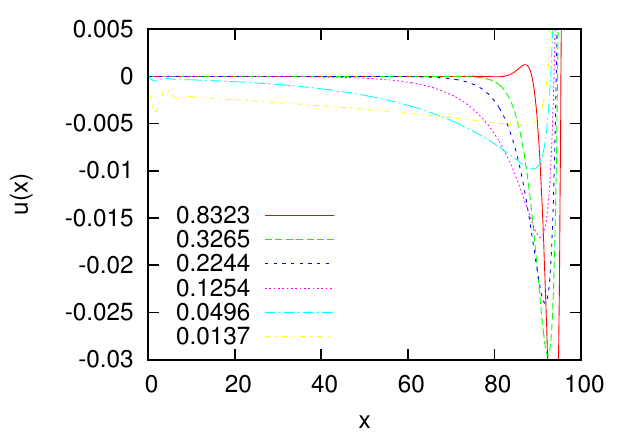}
    \caption{
        (Color online)
        The left tails of the solutions obtained while the tracing (b) to (c)
        whose $A_f$ is $0.8323$, $0.3265$, $0.2244$, $0.1254$, $0.0496$
        and $0.0137$ respectively.
        As $A_f$ decreases,
        the characteristic length of the tail becomes longer and longer.
        Finally an oscillation appears around the left boundary
        due to the boundary condition.
    }
    \label{fig:weakening}
\end{figure}

In the step (i), we fix $c$ to $1.21$ and execute a continuation with $A_f$.
This continuation leads $A_f$ around $0.01$, but $A_f$ cannot reach 0.
In this continuation the profile of the solution changes as shown in \cref{fig:weakening}.
We define a characteristic length of the tail of the solution
as the inverse of its decaying rate. 
As $A_f$ decreases, it becomes longer and an oscillation starts to appear
around the left boundary.
If this tracing is continued further, the oscillation grows up 
and the solution may not be kept localized.
Such non-localized solutions also cannot be
 traced till $A_f=0$.
This shows that a localized TWS to KSE cannot be obtained
 using only the continuation with $A_f$.

In order to avoid the oscillation around the boundary,
we focus on the tail of the solution.
In \cref{fig:weakening},
the oscillation seems to appear when the tail loses its flat part.
Indeed, at $A_f = 0.2244$ where its flat part is around $[0,60]$
and $A_f = 0.1254$ where it is around $[0,40]$ the oscillation does not appear.
We conclude that the disappearance of the flat part, which may occur when $A_f < 1.0$,
is a precursor of the oscillation at the boundary. 
We call this decrease of the flat part ``the weakening of the localization'',
and will discuss this mechanism in the next section.

\begin{figure}[tbp]
    \centering
    \includegraphics{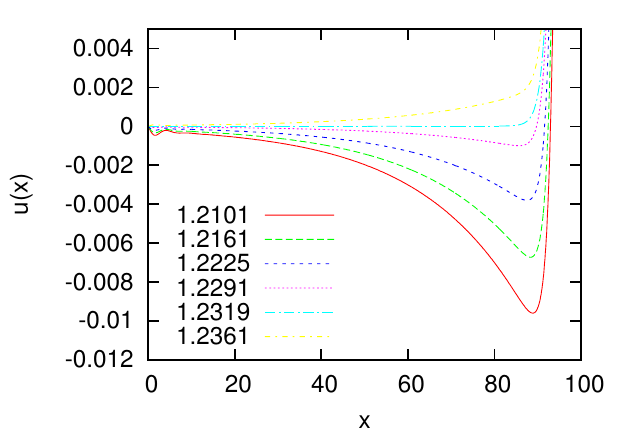}
    \caption{
        (Color online)
        A part of the profile of solutions obtained while the tracing (c) $\to$ (d)
        whose traveling velocity $c$ is
        $1.2101\text{(c)},1.2161,1.2225,1.2291,1.2319\text{(d)},1.2361$.
        The mechanism of this weakening of the localization
        is argued in \cref{sec:KSE3}.
    }
    \label{fig:wide_skirt}
\end{figure}

We found that
there exists a specific value $c$ for each $A_f$
such that the flat part recovers.
Figure \ref{fig:wide_skirt} shows the change in the tail of the solution
while the continuation with $c$ ($A_f$ is fixed to $0.0496$).
At $c=1.2319$ the flat part recovers to be $[0,80]$.
This continuation is the step (ii),
and the ``appropriate'' value is $c=1.2319$.

\begin{figure}[tbp]
    \centering
    \includegraphics{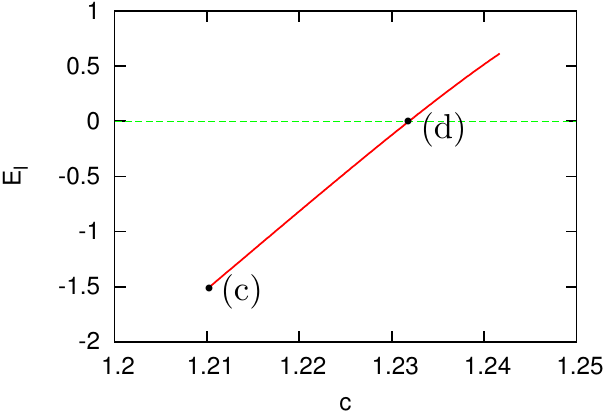}
    \caption{
        (Color online)
        $E_l$ is plotted against the traveling velocity $c$.
        $E_l$ measures how long the tail of the solution is.
        $E_l(c)$ cross the line $E_l = 0$,
        and this point is labeled (d).
        This crossing behavior is also argued in \cref{sec:KSE3}.
    }
    \label{fig:EdgeIntegral}
\end{figure}

\begin{figure}[tbp]
    \centering
    \includegraphics{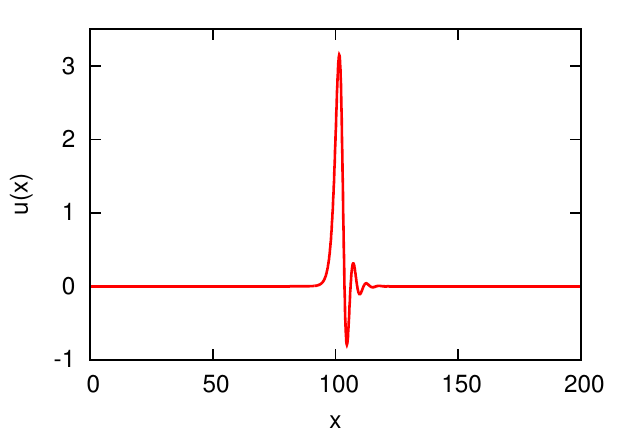}
    \caption{
        (Color online)
        The profiles of TWS to \cref{eq:KSE}.
        It takes non-zero value on a region $[90:115]$,
        which is much wider than $\Omega = [97:103]$.
    }
    \label{fig:KSE_last}
\end{figure}

This ``appropriate'' value of $c$ varies with $A_f$.
In other words, the ``appropriate'' relation  $c=c(A_f)$ defines
a path on the two-dimensional parameter space $c$-$A_f$.
Along this path the solution has always a flat part.
In order to trace the solution along the path $c=c(A_f)$,
however, it is necessary to express the condition $c=c(A_f)$ numerically.
It can be easily done with the following integral value:
\begin{equation}
    E_l = \int_0^l u(x) dx,
    \label{eq:EdgeIntegral}
\end{equation}
where $l$ is chosen to be in the tail part.
The change of $E_l$ in the  continuation mentioned above 
 is shown in \cref{fig:EdgeIntegral}.
The relation $c = c(A_f)$ is now implicitly defined by $E_l(c,A_f) = 0$.
Then we can continuate the branch $u(x; c(A_f), A_f)$ with $A_f$.
This conditional continuation can be implemented by a $(N+1)$-dimensional arc-length method,
and we have succeeded to trace the solution till $A_f=0$ within a numerical accuracy.
This solutions is labeled as (e) in \cref{fig:kse_tw_trace},
and its profile is displayed in \cref{fig:KSE_last}.
Its propagation velocity $c_0 = c(0)$ equals to $1.2143$.
This is the same solitary-wave solution
as that shown in fig.4c of \cite{Michelson1986}.

\subsection{Why the tail of the solution becomes longer?}
\label{sec:KSE3}
In this subsection the mechanism of the weakening of the localization is considered.
This weakening behavior enable us
to obtain an implicitly defined path in $A_f$-$c$ space.
In order to generalize this technique
to more complicated systems such as channel flows or pipe flows,
it is necessary to investigate its details more precisely.

A steady localized solution $u(x)$ to \cref{eq:KSEtw}
satisfies the ordinary differential equation:
\begin{equation}
    \frac{d^4 u}{dx^4} + \frac{d^2 u}{dx^2} + (u-c)\frac{du}{dx} + A_f H(x) u = 0.
    \label{eq:KSEtail}
\end{equation}
Here hats and primes are omitted for convenience.
Regarding $x$ as a virtual time,
this equation defines a four-dimensional dynamical system.
Then a localized solution $u(x)$ to \cref{eq:KSEtw} corresponds
to a homoclinic orbit connecting the saddle point of this dynamical system,
and the tails of the solution describes the asymptotic behavior of the homoclinic orbit
from and to the saddle point.
It should be noted that
this reinterpretation has less compatibility
with the fixed boundary condition $u(0) = u(L) = 0$
since we now consider the asymptotic behaviors in an infinite region $(-\infty, \infty)$
instead of the bounded region $[0,L]$.
However the following arguments are all valid whenever $u(x)$ decays fast enough
to be negligible comparing with the truncation error.

This dynamical system has a trivial fixed point (saddle point)
$(u,\partial_x u, \partial_{xx} u, \partial_{xxx} u) = (0,0,0,0)$
corresponding to the solution $u(x) = 0$ to \cref{eq:KSEtw}.
The localized solution $u(x)$ can be regarded as a homoclinic trajectory
of this trivial fixed point.
Then the tail of the localized solution can be analyzed by the eigenvalues
of Jacobi matrix of this dynamical system at the fixed point.
Since $H(x)$ equals to $1$ in the tail region,
the eigen equation of the Jacobi matrix becomes as follows:
\begin{equation}
    \lambda^4 + \lambda^2 - c\lambda + A_f = 0.
    \label{eq:EigenEquation}
\end{equation}
This quartic equation has two real roots $\lambda_0, \lambda_1$
and two complex roots $\lambda_\pm$.
Since we focus on the case $A_f \ll 1$,
we first consider the case $A_f=0$, and then a perturbation expansion about $A_f$.

When $A_f = 0$, the eigenvalues are $\lambda_0 = 0$ and
three roots of a cubic equation $\lambda^3 + \lambda -c = 0$.
This cubic equation has a real root $\lambda_1$ and two complex roots
$\lambda_\pm = (- \lambda_1 \pm i \sqrt{3\lambda_1^2 + 4})/2$.
The real nonzero root $\lambda_1$ is positive when $c>0$ and negative when $c<0$,
and here we consider the $c>0$ case.
Then the left and right tails of the solution can be written as follows:
\begin{align}
    u_L(x) &= A_1 e^{\lambda_1 x} \label{eq:KSE_tail0L} ,\\
    u_R(x) &= A_+ e^{\lambda_+ x} + A_- e^{\lambda_- x}  \label{eq:KSE_tail0R}.
\end{align}
The coefficients $A_1, A_+, A_-$ are determined in the nonlinear region.
Since  $\exp(\lambda_0 x)$ does not goes to zero as $x \to \pm \infty$,
$u(x)$ cannot contain this term.

When $0 < A_f \ll 1$ the zero eigenvalue is modified
as $\lambda_0 = A_f / c + O(A_f^2)$.
Then the left tail of the solution can be written as follows:
\begin{equation}
    u_L(x) = A_0 e^{\lambda_0 x} + A_1 e^{\lambda_1 x}.
    \label{eq:KSE_tail1}
\end{equation}
The coefficients $A_0$ and $A_1$ are also determined in the nonlinear region,
and depend both on $c$ and $A_f$.
Each of the terms in \cref{eq:KSE_tail1} defines a tail whose
characteristic length is $1/\lambda_0$ and $1/\lambda_1$,
and the realized tail is a superposition of them.
As $A_f$ goes to zero, the characteristic length $1/\lambda_0 = c/ A_f$ diverges.
Thus the modified eigenvalue $\lambda_0$ is the origin of the weakening, i.e., the long tail.

Next, we consider why we can obtain a solution with a short tail
by the condition $E_l = 0$ for every small $A_f$.
Using \cref{eq:KSE_tail1}, $E_l$ can be written as follows:
\begin{equation}
    E_l \simeq \int_{-\infty}^l u_L(x) dx
    = \frac{A_0}{\lambda_0}e^{\lambda_0l} + \frac{A_1}{\lambda_1}e^{\lambda_1l}.
    \label{eq:EdgeIntegralTail}
\end{equation}
Since $1/\lambda_0 \gg 1/\lambda_1$ when $A_f \ll 1$,
the first term in \cref{eq:KSE_tail1} is dominant
except for a region near the nonlinear region.
Thus the first term in \cref{eq:EdgeIntegralTail} is rather dominant
for an appropriate $l$.
Then $E_l$ roughly measures $A_0$,
and $E_l = 0$ is realized when $A_0$ is zero
where the tail by $\lambda_0$ disappears.
A more precise argument is also possible.
The condition $E_l = 0$ can 
yield the condition of $A_0$ as follows:
\begin{equation}
    \begin{split}
        A_0 &= - \frac{\lambda_0}{\lambda_1}A_1 e^{(\lambda_1 - \lambda_0)l} \\
            &= - \frac{A_f}{c\lambda_1}A_1 e^{(\lambda_1 - \lambda_0)l} + O(A_f^2).
    \end{split}
    \label{eq:ConditionA0}
\end{equation}
Thus $A_0$ is not exactly zero while $A_f > 0$.
However, since $\lambda_0$ goes to zero as $A_f \to 0$,
$A_0$ satisfying this condition also goes to zero.

The essence of the above arguments is
the existence of the zero eigenvalue $\lambda_0$
and its modification due to the damping.
The modified zero eigenvalue introduces another degree of freedom
in the determination of tails of solutions.
Calculating the eigenvector of the zero eigenvalue,
it corresponds to a uniform level raising of the velocity field,
$u(x) \mapsto u(x) + \delta c$,
so it corresponds to Galilean invariance.
In other words, the reason why this small eigenvalue appears
is the breakdown of Galilean invariance.
This fact indicates that such the 
weakening of the localization is expected to occur
whenever a stream-wisely localized TWS is going to be obtained
in Galilean-invariant systems by our method.

At last, we conclude this section with an error estimate.
Since the continuous parameter dependence on $c$ disappears when $A_f = 0$,
the continuation becomes unstable as $A_f$ goes to zero.
Although the point $A_f = 0$ is a singular point in this continuation problem,
$A_f$ can get an arbitrary value as small as the numerical accuracy allows.
Indeed, we get $A_f \sim 10^{-10}$ in the conditional continuation.
This is as small as a threshold for Newton method, $\varepsilon_{\text{Newton}}$.
Then the continuated solution can be regarded as a solution to KSE
within an numerical error $\varepsilon_{\text{Newton}} + A_f \| u\|$.

\section{\label{sec:ConcludingRemarks}Concluding Remarks}
In this paper we introduce the damping filter method
for obtaining spatially localized solutions.
We adopt our method into two fundamental cases.
First, in the \cref{sec:SHE},
we consider localized solutions to Swift-Hohenberg equation (SHE).
Then our method can not only reproduce known solutions,
but also obtain another spatially localized solution
which belongs to a closed isolated solution branch.
Next, in the \cref{sec:KSE},
we consider a stream-wisely localized traveling-wave solution (TWS)
to Kuramoto-Sivashinsky equation (KSE).
In this case, since the propagation velocity $c$ is unknown,
we have to continuate the solution with $c$ and the filter amplitude $A_f$.
In order to make continuation one-dimensional we introduce an implicit condition
about the tail of solutions.
Here we reinterpret these result from a general point of view
in order to adopt our method into more general cases.

The most interesting result in the \cref{sec:SHE} is the connection
between two distinct solution branches.
We show that two solution branches are connected with each other
through the continuation with the filter amplitude $A_f$.
We first introduce the filter term $-A_f H(x) u$
in order to obtain a guess at spatially localized solutions.
However, it works as an excitation when $A_f < 0$.
This excitation only works near the localized region
where both $u(x)$ and $H(x)$ are non-zero.
This causes an instability which may lead another localized solution.
Indeed, we obtain the solution (d) in \cref{sec:SHE} only by the continuation with $A_f$.
This spatially selective excitation mechanism has an advantage
in searching spatially localized solutions.
The result for SHE indicates that if another localized solution exists
near the localized solution already obtained in the phase space,
they may connect through this excitation mechanism.
So our method may enable us to search localized solutions automatically.

In the \cref{sec:KSE},
we have dealt with a spatially localized TWS to KSE.
The main issue of this section is a treatment of the invariances. 
The damping filter term breaks the translational and Galilean invariances, 
but they recover when the filter disappears.
This singular behavior is avoided by imposing
the condition $c = c(A_f)$ by the implicit condition $E_l(c,A_f) = 0$.

This artificial condition can be reinterpreted as a critical line of 
the orbit-flip bifurcation \cite{CHAMPNEYS1996}.
For fixed $A_f$ an orbit-flip bifurcation occurs when $A_0$,
defined in \cref{sec:KSE3}, changes its sign with increasing $c$,
see \cref{fig:RestrictedPath} and \cref{fig:wide_skirt}.
Moreover, if $A_f \ll 1$, the relation $A_0 \sim A_f \sim 0$ holds
on $c = c(A_f)$ because of \cref{eq:ConditionA0}.
Therefore we can infer that the condition $c = c(A_f)$ corresponds
to the critical line of the orbit-flip bifurcation
in the four-dimensional ODE system \cref{eq:KSEtail}.

To implement the critical condition directly,
we can utilize an algorithm for tracking the orbit-flip bifurcation in AUTO \cite{doedelauto}.
This algorithm replaces the condition $E_l = 0$ with an orthogonal condition
to keep the tangency between the homoclinic orbit and the leading eigenspace.
Moreover, this method clarifies the mathematical meaning of our condition.
However, since it is designed for a homoclinic orbit to ODE,
it might be difficult to apply it for spatially two- or three-dimensional PDE systems. 
We thus expect that our method using $E_l$ will be more suitable
for the dynamical systems approach to turbulence.

\begin{acknowledgments}
We thank T.Ogawa for invaluable comments on Swift-Hohenberg equation.
We also appreciate the referee referring us to the orbit-flip bifurcation.
This work was supported by the Grants for Excellent Graduate Schools
``The Next Generation of Physics, Spun from Universality and Emergence''
from the Ministry of Education, Culture, Sports, Science, and Technology (MEXT) of Japan,
and also partially by JSPS KAKENHI Grant Number 22540386.
A part of numerical calculations were carried out on SR16000 at YITP in Kyoto University.
\end{acknowledgments}

\bibliography{paper1}
\end{document}